\documentclass[10pt,journal,compsoc]{IEEEtran}
\ifCLASSOPTIONcompsoc
  \usepackage[nocompress]{cite}
\else
  \usepackage{cite}
\fi
\ifCLASSINFOpdf
\else
\fi
\usepackage[utf8]{inputenc}
\usepackage{amsfonts}
\usepackage{amsmath}
\usepackage{bm}
\usepackage{float}
\usepackage{booktabs}
\usepackage{stfloats}
\usepackage{diagbox}
\usepackage{amsthm}
\usepackage{pgfplots}
\usepackage{tikz}
\newtheorem{Def}{Definition}
\newtheorem{Theo}{Theorem}
\allowdisplaybreaks[4]
\usepackage{multirow}
\usepackage{color}
\usepackage{graphicx}
\usepackage[english]{babel}

\begin{document}
	\newcommand\wh{\widehat}
	\newtheorem{remark}{Remark}
\title{Two-Server Delegation of Computation on Label-Encrypted Data}

\author{Xin~Chen, Liang Feng~Zhang
\IEEEcompsocitemizethanks{\IEEEcompsocthanksitem Xin Chen and Liang Feng Zhang are with School of Information Science and Technology,  ShanghaiTech University, Shanghai, China. (E-mail: chenxin3@shanghaitech.edu.cn, zhanglf@shanghaitech.edu.cn)}}

\markboth{Journal of \LaTeX\ Class Files,~Vol.~14, No.~8, August~2015}%
{Shell \MakeLowercase{ {et al.}}: Bare Demo of IEEEtran.cls for Computer Society Journals}

\IEEEtitleabstractindextext{
\begin{abstract}
	Catalano and Fiore propose a scheme to transform a linearly-homomorphic encryption into a homomorphic encryption scheme capable of evaluating quadratic computations on ciphertexts. Their scheme is based on the linearly-homomorphic encryption (such as Goldwasser-Micali, Paillier and ElGamal) and need to perform large integer operation on servers. Then, their scheme  have numerous computations on the servers. At the same time, their scheme cannot verify the computations and cannot evaluate more than degree-4 computations. To solve these problems, we no longer use linearly-homomorphic encryption which based on number theory assumptions. We use label and pseudorandom function to encrypt message, which significantly reduce the computations on the servers and enable us to use homomorphic MACs technology to realize verifiable computations naturally. We also extend the method to construct $d$-server schemes, which
 allow the client to delegate  degree-$d$ computations on outsourced data.
\end{abstract}

}

\maketitle
\IEEEdisplaynontitleabstractindextext
\IEEEpeerreviewmaketitle

\IEEEraisesectionheading{\section{Introduction}\label{sec:introduction}}

\IEEEPARstart{T}{he} prevalence of cloud computing makes it  very popular for the client such as the users of resource-restricted devices to collect data,  outsource the data to one or more cloud services, and
later freely access  the data on demand,  even if the client have
very limited storage   or computing power.
The client may not only access the outsourced data by retrieving one or more specific elements, but also
request the cloud services to perform  computations on the outsourced data and
then return the correct results.
If there is only one cloud service, then the   outsourcing  scenario above can be described as follows. A client
collects a set of data elements $m_1,\ldots,m_n$, stores  these elements
on a cloud server, and later asks the server to run a program $\cal P$ over
$(m_1,\ldots,m_n)$.    The server computes $m={\cal P}(m_1,\ldots,m_n)$ and returns
$m$.

This simple scenario has incurred significant security concerns.
The  attacks \cite{att}  show that one cannot always   trust the cloud services by storing sensitive information
on their servers, as the cloud services may not be able to  always
defeat the attackers from both inside and outside.
How to preserve the {\em privacy} of the outsourced data is one of the top security concerns. Encrypting the data with the traditional   algorithms such as AES \cite{aes} and RSA \cite{rsa} would
not only  allow the client to preserve the data privacy  but also
make the server-side computation of ${\cal P}(m_1,\ldots,m_n)$ impossible.
A natural way to resolve this technical difficulty is to use the
homomorphic encryption schemes.
Fully homomorphic encryption scheme (FHE) \cite{fhe} allows
the server to perform the computation of any program $\cal P$
on the ciphertexts ${\sf Enc}(m_1),\ldots, {\sf Enc}(m_n)$, instead of the plaintexts,
to get a ciphertext  of  $m={\cal P}(m_1, \ldots,m_n)$.
The invention of FHE \cite{fhe} has been a main breakthrough in
  cryptography.  However, today's FHE constructions \cite{GSW13,BGV12,BV11,BV14} still suffer  from
large parameters and  are rather slow. As a result, the FHE-based outsourcing    is  time-consuming and far from practical.

The notion of homomorphic encryption dates back to
Rivest, Adleman and Dertouzous \cite{RAD78}.
The first homomorphic encryption schemes were  constructed
by \cite{GM84,Pai99}.  On one hand, these schemes   allow only linear computations on the encrypted data.
On the other hand, these schemes are much more efficient
than FHE  \cite{fhe,GSW13,BGV12,BV11,BV14}.
In the outsourcing computation scenario, if the client is only interested in a linear combination of
the outsourced data, then the FHE   can be replaced with a  linearly homomorphic encryption scheme
\cite{GM84,Pai99} and  results in strictly faster schemes.

It is   possible to extend the linearly homomorphic encryption schemes to a new encryption scheme that
enables the computation of nonlinear functions on the ciphertexts.
Catalano and Fiore \cite{CF14}
proposed a transformation that can
convert any public-space linearly-homomorphic encryption scheme (the message space is a publicly known ring) into a homomorphic encryption scheme supporting
quadratic computations. The outcome scheme of their transformation would allow
quadratic computations in the outsourcing scenario, but
with a blow-up of ciphertexts.
Based on the transformation, they constructed a two-server scheme for
delegating quadratic computations on  the outourced data, where the blow-up in ciphertext/communication is
avoided in a clever way.
In their scheme, each data element $m_i$ is encrypted as a pair $(m_i-a_i,{\sf Enc}(a_i))$ and given to the first server
and the random number $a_i$ is given to the second server, where $\sf Enc$ is any linearly homomorphic encryption scheme. The computation of $m_1m_2$ is done by the first server
computing a ciphertext $c={\sf Enc}((m_1-a_1)(m_2-a_2)+a_1(m_2-a_2)+a_2(m_1-a_1))$
and the second server computing $a_1a_2$.  The client
learns $m_1m_2$ by computing ${\sf Dec}(c)+a_1a_2$.
The privacy of data is achieved by assuming that
$\sf Enc$ is semantically secure and the two servers do not collude with each other.

While Catalano and Fiore's two-server scheme \cite{CF14} allows one to delegate quadratic computations
using  LHE and in a succinct manner,  the server-side computations however can be
slow provided that the number $n$ of data elements  $m_1,\ldots,m_n$ is large.
{For example, when $n=10^3$ and the $\sf Enc$ is chosen as the fast Paillier's encryption  \cite{Pai} the server-side computations
for a quadratic function may require as much as
191  seconds.} The waiting time could be a main measure of the clouds' service quality and
a poor quality would discourage the client from actually using the service.

Neither  could the client trust the cloud services by simply storing sensitive information in clear
on their servers, nor the client could trust these services by simply accepting their computation results.
After all, the cloud services have the financial incentive to run an extremely fast
but incorrect computation, in order to free  up valuable computing time for other transactions.
 How to enforce the {\em integrity} of the server-side  computations
	is also among the top  security concerns in    outsourcing computation.
The problem of enforcing server-side computations' integrity has been extensively studied
under the name of securely outsourcing computation and realized with
verifiable computation \cite{GGP10,CKV10,BGV11,FG12}, homomorphic message authenticators \cite{GW13,GVW15,CF18}, and many other
primitives \cite{BCCT12,GW11,GGPR13,PHGR16}.
In Catalano and Fiore's two-server scheme \cite{CF14}, each server is completely trusted
  to perform the specified computations correctly.
  However, a dishonest server may easily change the client's output by sending back
an arbitrarily chosen result (an LHE ciphertext in the first server and a ring element in the second server).

We consider the long waiting time of service and the
lack of integrity of server-side computations as two main drawbacks of Catalano and Fiore \cite{CF14}.
It is an interesting   problem to devise delegation of computation schemes with both {\em practically  fast} server-side computations and  {\em integrity} of server-side computations.

\subsection{Our Contributions}
In this paper, we introduce a  model called
{\em two-server delegation of computation on label-encrypted data} ({\sf 2S-DCLED}),
 in order to provide a solution to the problem as above in the scenario of outsourcing
 computations. The idea of associating data elements with labels has been used in
 \cite{BCF17} to build a {\em labeled homomorphic encryption}, which supports the
 quadratic homomorphic computations on ciphertexts and resolves the
  compactness issues of \cite{CF14}. Our model is obtained by integrating
   this idea into the two-server delegation of computation on encrypted
    data model of \cite{CF14}. In the new {\sf 2S-DCLED} model we proposed
     two schemes for delegating quadratic computations of the outsourced data
      on two non-communicating servers. Both schemes keep the client's data private
       from each individual server under the mild assumption that PRFs exist.
        Comparing with \cite{CF14},  the server-side computations  in our first
         scheme is $\geq$ 2200 times faster, which significantly reduces the
         waiting time of service. Our second scheme adds integrity of server-
         side computations to the first scheme by using the homomorphic MAC
         of  \cite{CF18}, at the price of slightly slowing the server-side
         computations. For every integer $d>2$, we also extend the model
         of {\sf 2S-DCLED} to   the model of {\em $d$-server delegation of
         computation on label-encrypted data} ($d${\sf S-DCLED}). We  devise
           $d${\sf S-DCLED} schemes that enable the delegation of degree-$d$
            computations on the outsourced data by using $d$ non-communicating
             servers.

\subsection{Our Techniques}

Our design starts from accelerating the server-side computation in
the {\sf 2S-DCED} scheme of \cite{CF14}. Our implementation of the scheme    \cite{CF14}
shows that the most time-consuming part of the server-side computations
in \cite{CF14}  consists of  the homomorphic computations over
$\{(m_i-a_i,{\sf Enc}(a_i))\}_{i=1}^n$, which are done by the first server and require  a large amount of public-key operations
such as exponentiations modulo a large integer.
Our basic idea of accelerating sever-side computations is based on removing
  the dependence on $\sf Enc$, the linearly homomorphic encryption scheme.
In   \cite{CF14} any quadratic computation  of the form $m_1m_2$
was decomposed as
$$m_1m_2=(m_1-a_1)(m_2-a_2)+a_1(m_2-a_2)+a_2(m_1-a_1)$$
$$\hspace{-4.5cm}+a_1a_2.$$
 The first server is given $(m_1-a_1,{\sf Enc}(a_1)),(m_2-a_2,{\sf Enc}(a_2))$
and   responsible to compute a ciphertext $c$ of
$(m_1-a_1)(m_2-a_2)+a_1(m_2-a_2)+a_2(m_1-a_1)$; the second server is given
$a_1,a_2$ and responsible to compute $a_1a_2$.
The reconstruction is done by computing ${\sf Dec}(c)+a_1a_2$.
The privacy of data is based on the assumption that
$\sf Enc$ is semantically secure and the two servers do not collude with each other.

In our design, the $\sf Enc$ will be removed in order to accelerate the
server-side computations. As a consequence, the
first sever is no longer able to include $a_1(m_2-a_2)+a_2(m_1-a_1)$ in  $c$
and the client will not be able to recover
$m_1m_2$ merely   from $c={\sf Enc}((m_1-a_1)(m_2-a_2))$ and $a_1a_2$, as $m_1, m_2$ are both unknown to
the client.
To bypass this technical difficulty, we
offload the linear computations such as   $a_1m_2+a_2m_1$ to the second server
such that together the results from both servers would enable the client to
remove these terms.
Our key observation  is a new decomposition of quadratic computations of the form
$m_1m_2$ as below
\begin{align*}
m_1m_2=&(m_1-a_1)(m_2-a_2)-(a_1-b_1)(a_2-b_2)\\
&+a_1(m_2-b_2)+a_2(m_1-b_1)+b_1b_2,
\end{align*}
where $a_1,a_2,b_1,b_2$ can be any elements from the domain of data elements.
This decomposition allows us to preserve the privacy of
$m_1,m_2$ against each individual server by sending
$(m_1-a_1,a_1-b_1), (m_2-a_2,a_2-b_2)$ to the first server and
sending $(m_1-b_1,a_1),(m_2-b_2,a_2)$ to the second server.
If we instruct the first server to compute
$c_1=(m_1-a_1)(m_2-a_2)-(a_1-b_1)(a_2-b_2)$ and the second server to compute
$c_2=a_1(m_2-b_2)+a_2(m_1-b_1)$, then
the value of $m_1m_2$ would be easily recovered as $c_1+c_2+b_1b_2$, where  $b_1b_2$ can be computed
on the client's local devices.
Quite different from \cite{CF14}, the privacy of data in our design is
based on the  assumption that the two servers do not collude.
This is because the linearly homomorphic encryption $\sf Enc$ is no longer used.
The server-side computation will be significantly accelerated as well since
no public-key operations are involved.
These improvements are not obtained at no price.
In order to recover $m_1m_2$, the client in our design has to
compute $b_1b_2$ locally, which will slow the client-side computation.
Nevertheless we shall show with experiments that for
moderately large data set size $n$, the client-side computing cost in our design is still
much lower than \cite{CF14}.
Our client has to remember the  random numbers
$b_1,b_2,\ldots,b_n$.
While this is not very  satisfactory, especially when the client's device has very limited
storage capacity, we deal with the difficulty by associating each data element
$m_i$ with a label $\tau_i$ and generate $b_i$ as a pseudorandom value  $F_K(\tau_i)$, where
$F$ is a PRF.
In such as way, we obtained a {\sf 2S-DCLED} scheme where the data privacy is simply based on the mild
assumption that PRFs exist and the two servers do not collude,   the server-side computations
are significantly faster, and the client-side computations are also faster than \cite{CF14} when the size of data is moderate.
Although the client-side computation is even slower than the delegated
 computation, the {\sf 2S-DCLED} is still meaningful as long as the client
  is short of storage. In fact, our schemes in this paper will be
  specifically designed for the storage-restricted devices.

Note that in our {\sf 2S-DCLED} scheme the server-side computations have very good forms. In fact, both
servers only need to perform polynomial computations of degree 2 on their stored data.
More precisely, when the message space is $\mathbb{Z}_p$, the finite field of $p$ elements for a prime $p$,
Catalano and Fiore \cite{CF18} has proposed a homomorphic MAC that enables the client to authenticate
the data elements  $m_1,\ldots,m_n$ with tags $t_1,\ldots,t_n$ such that any polynomial computation over
$m_1,\ldots,m_n$ can be authenticated with a similar computation over
the tags $t_1,\ldots,t_n$.
Let $f:\mathbb{Z}_p^n\rightarrow \mathbb{Z}_p$ be any polynomial function.
In the scheme of \cite{CF18}, both a random field element $s$ and a key $K$ for PRF
$F$ are chosen as the secret   key; each data element $m_i$ is authenticated with
a tag $t_i(x)=m_i+\frac{F_k(\tau_i)-m_i}{s}x$, a univariate polynomial over $\mathbb{Z}_p$
such that $t_i(0)=m_i$ and $t_i(s)=F_K(\tau_i)$.
In order to learn $f(m_1,\ldots,m_n)$, the client simply gives $f$ to the server,
the server sends back $y=f(m_1,\ldots,m_n)$ and $t(x)=f(t_1(x),\ldots,t_n(x))$.
The client accepts $y$ if and only if $t(0)=y$ and $t(s)=f(F_K(\tau_1),\ldots,F_k(\tau_n))$.
The server-side computations in our design are   quadratic polynomial computations on the stored data.
By applying the homomorphic MAC of \cite{CF18} between the client and each individual server, we
are able to add computation integrity to the
{\sf 2S-DCLED} scheme and obtain a scheme with with computation integrity, which is called 2S-VDCLED.
  While the data privacy is not changed, the computation integrity
is also based on the sole assumption that PRFs exist.
The additional price of adding integrity is that the client-side computation will be slightly slowed.
However, we stress  that, for moderately large data set size, our client-side computations is still much faster  than \cite{CF14},
a scheme without integrity.

\subsection{Evaluations and Comparisons}

Both the {\sf 2S-DCED} scheme of \cite{CF14} and our {\sf 2S-DCLED}/{\sf 2S-VDCLED} schemes allow the
delegation of quadratic computations over outsourced data.
In \cite{CF14} the  privacy of data is based on the assumption that
the underlying LHE is semantically secure and the two servers do not collude.
Our data privacy is  based on the weaker assumption that PRFs exist and the two servers do not collude.
While the {\sf 2S-DCED} and {\sf 2S-DCLED} schemes provide no computation integrity,
our {\sf2S-VDCLED} scheme can prevent the client from accepting a wrong result.
While the servers of   \cite{CF14} have to do a large number of  exponentiations   modulo  large integers,
our server-side computations   only involve multiplications over much smaller fields and
are much faster.
In terms of client-side computations, although our schemes are not asymptotically better,
they are still much faster when the size of data is moderate.

It is  possible to extend \cite{CF14} to support  the computation of  degree-3 polynomials. However,  the server-side computational cost   will increase sharply. One cannot use \textsf{2S-DCED} to evaluate
functions of degree $\geq 4$. In contrast,  by using more servers we can outsource
 the computation of  functions of
arbitrary  degrees.

\begin{figure*}[htp]
	\centering
	\includegraphics[width=0.8\linewidth]{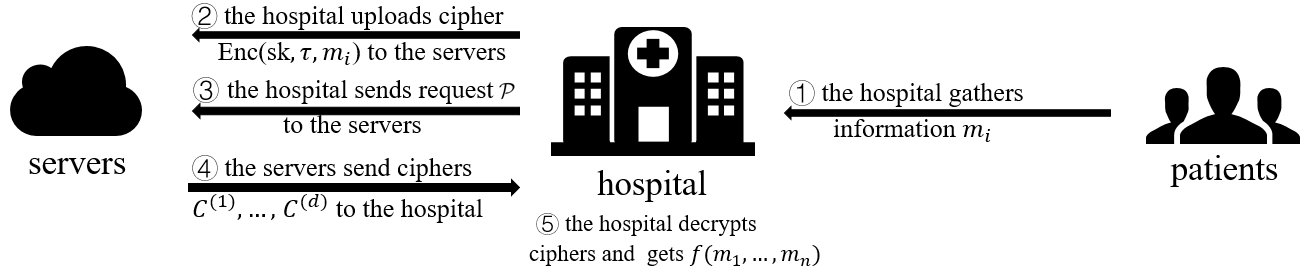}
	\caption{Disease Diagnosis}
\end{figure*}

\begin{figure*}[htp]
	\centering
	\includegraphics[width=0.8\linewidth]{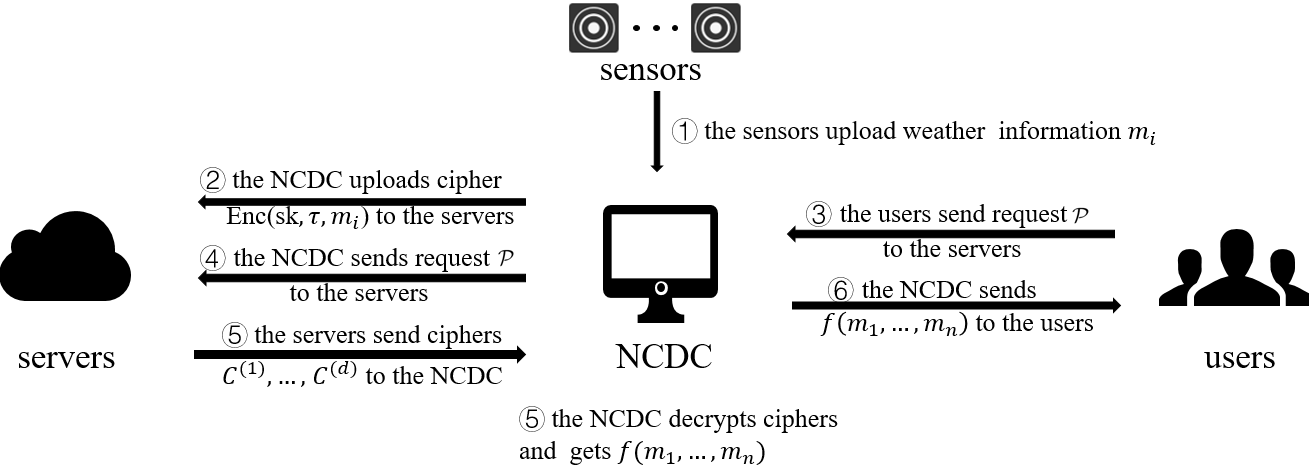}
	\caption{Data Analysis}
\end{figure*}
\subsection{Related Works}

Barbosa  {et al.} \cite{BCF17} constructed a   {labeled homomorphic encryption} scheme which  allows a client to delegate quadratic computations on the outsourced encrypted  data to a
{\em single} server. Their schemes neither  support verification of servers' computations nor
allow   degree-3 computations on the outsourced data.
Catalano and Fiore \cite{CF18} proposed a homomorphic MAC scheme
that enables the client to authenticate
the data elements  $m_1,\ldots,m_n$ with tags $t_1,\ldots,t_n$ such that any polynomial computation over
$m_1,\ldots,m_n$ can be authenticated with a similar computation over
the tags $t_1,\ldots,t_n$.
Their schemes cannot keep the outsourced data private.
Zhang  {et al.} \cite{ZS14} proposed a  verifiable local computation  model where
the client can privately outsource data elements to cloud servers and later verify
computations on any portion of the outsourced data.
Their schemes satisfy our security and efficiency requirements, but require  at least $d+1$
non-communicating servers in order to compute degree-$d$ functions.
Tran  {et al.} \cite{TPD16} proposed two single-server schemes based on homomorphic MACs and do not
rely on FHE. One of their schemes
supports the computation of  quadratic functions on outsourced data.
Unfortunately,
it  has been broken    \cite{XHZ17}.

\subsection{Application}

\noindent
{\bf Disease Diagnosis.}
In this application, a hospital records
the patient's physical examination data for the doctor to diagnose.
However, storing data will consume a lot of
resources. It is difficult for a hospital to have a machine
that can process function on numerous data. Then, it is
feasible   for hospital to store  the data on cloud
servers and perform desired computations on the outsourced data.
Our scheme can solve these problems, and the Fig. 2  shows
the process.

\vspace{1mm}
\noindent
{\bf Data Analysis.}
National Climatic Data Center(NCDC) has many sensors for temperature
and pressure, which are distributed throughout the world. The sensor transmits the data to the NCDC at various times.
The NCDC stores the data for sale to other users who are individuals
or institutions that need to use the data for research. In this scenario,
there are big flaws in data collection and sales. NCDC stored at
least 13.4 PB of data whose maintenance will consume a lot of
  resources. The users need to download all
the data used in the research, and the communication cost is high. In
order to solve these problems, we propose a new scheme whose core idea
is to encrypt the data and outsource it to the server provider. The
Fig. 2 presents problems solving process.

\vspace{1mm}
\noindent
{\bf Neural Networks.}
Applying neural networks to a problem which involves sensitive data requires accurate predictions and maintaining data privacy and security. \cite{GRNK} have solve this problem, they approximate these non-linear functions using low-degree polynomials, such that the modified neural network can be evaluated using a FHE\cite{fhe}. Compared to FHE, our scheme evaluate neural networks over private data more efficiently.

\vspace{1mm}
\noindent
{\bf Moments.}
In mechanics and statistics, a moment is a specific quantitative measure
of the shape of a function. Since the $d$-th moment is computable by a
degree-$d$ polynomials, our scheme can compute the $d$-th moment by
$d$ servers.

\vspace{1mm}
\noindent
{\bf Polynomials  with Hidden Coefficients.}
The clients using Shamir secret sharing\cite{ss} can hide the coefficients
of the monomials in $f$, and turning $f$ into a degree-($d+1$) polynomial
$f'$. Furthermore, the clients can hide the monomials in $f$ by Shamir
secret sharing the coefficients in $f$ of all monomials of degree at
most $d$.

\subsection{Organization}

In Section 2 we formally define the model of two-server delegation of computation on label-encrypted data;
In Section 3 we present a specific construction of 2{\sf S-DCLED} scheme;
Section 4 contains a 2{\sf S-DCLED} scheme that also satisfies the
unforgeability property; In Section 5 we extend the model of 2{\sf S-DCLED}
to  $d${\sf S-DCLED} scheme for any integer $d\geq 2$.
In Section 6 we implement the 2-server schemes of Section 3 and 4, and
compare them with the 2-server schemes from \cite{CF14}.
Finally, Section 7 contains our concluding remarks.

\section{Preliminaries}
\textbf{Notation.}
We denote with $\lambda \in \mathbb{N}$ a security parameter, and with
 $\mathsf{poly}(\lambda)$ any function bounded by a polynomial in $\lambda$.
  We say that a function $\epsilon$ is  {negligible} if it vanishes
  faster than the inverse of any polynomial in $\lambda$. We use PPT for
  probabilistic polynomial time. If $S$ is a set, $x{\leftarrow}S$ denotes
  selecting $x$ uniformly at random from $S$. If $\mathcal{A}$ is a
  probabilistic algorithm, $x{\leftarrow}\mathcal{A}(\cdot)$ denotes the
   process of running $\mathcal{A}$ on some appropriate input and assigning
    its output to $x$. For a positive integer $n$, we denote by $[n]$ the
    set $\{1,\dots,n\}$.
    Let $X$, $Y$ be two random variables over a finite set $\mathcal{U}$. We define the statistical distance between $X$ and $Y$ as
	$$\mathsf{SD}[X,Y]=\frac{1}{2}\sum_{u\in U}|\Pr[X=u]-\Pr[Y=u]|.$$

\subsection{Labeled Programs}
A {\em labeled program} \cite{BCF17} $\mathcal{P}$ is a tuple  $(f,\tau_{1},\dots,\tau_{n})$ such that $f:\mathcal{M}^n\rightarrow \mathcal{M}$ is an $n$-ry function over the message space $\cal M$, and each label $\tau_i \in \{0,1\}^*$ uniquely identifies the $i$-th input of $f$. Composition of labeled programs works as follows. Given labeled programs $\mathcal{P}_1,\dots,\mathcal{P}_t$ and a function $g:\mathcal{M}^t\rightarrow \mathcal{M}$, the composed program $\mathcal{P}^*$ is obtained by evaluating $g$ on the outputs of $\mathcal{P}_1,\dots,\mathcal{P}_t$. Such a program is denoted as $\mathcal{P}^*=g(\mathcal{P}_1,\dots,\mathcal{P}_t)$. The (labeled) inputs of $\mathcal{P}^*$ are all of the distinct labeled inputs of $\mathcal{P}_1,\dots,\mathcal{P}_t$ (all inputs sharing the same label are
considered as   a single input to the new program). Let $f_{id}:\mathcal{M}\rightarrow\mathcal{M}$ be the canonical identity function and let $\tau\in\{0,1\}^*$ be a label. We denote by $\mathcal{L}_{\tau}=(f_{id},\tau)$ the  { identity program} for input label $\tau$. With this notation, any labeled program $\mathcal{P}=(f,\tau_1,\dots,\tau_n)$ can be expressed as the composition of $n$ identity programs, i.e., $\mathcal{P}=f(\mathcal{L}_{\tau_1},\dots,\mathcal{L}_{\tau_n})$.
\subsection{Two-Server Delegation of Computation on Label-Encrypted Data}

 A   two-server delegation of computation on label-encrypted data (\textsf{2S-DCLED}, for short) scheme
 is a communication protocol between a client and two non-communicating servers.
 It allows the client to encrypt any data  item  as two ciphertexts, one for each server, and then
 outsource the computation of a program to the servers.
Each server performs a computation  of the program on  its ciphertexts and
returns a partial result.
The client can  reconstruct the output of
the program.
The encryption should keep   each individual server  from learning any information about the data items.
 Formally, a
two-server delegation of computation on label-encrypted data
 scheme     {\sf 2S-DCLED}=($\mathsf{2S.KeyGen}$, $\mathsf{2S.Enc}$, $\mathsf{2S.Eval_1}$, $\mathsf{2S.Eval_2}$, $\mathsf{2S.Dec}$) consists of the following algorithms:

\vspace{2mm}

 $\mathsf{2S.KeyGen}(1^\lambda)$: This is a  key generation algorithm.
 It takes the security parameter $\lambda$ as input and produces  a secret key $\mathsf{sk}$ and a public key $\mathsf{pk}$.
	
 $\mathsf{2S.Enc(sk,\tau,m)}$: This is an   encryption algorithm.
 It takes the secret key  $\mathsf{sk}$,   any message $m\in {\cal M}$ and its label
 $\tau$ as input, and  outputs two ciphertexts $C^{(1)}$ and $C^{(2)}$.
	
$\mathsf{2S.Eval}_i(\mathsf{pk},\mathcal{P},C^{(i)}_1,\dots,C^{(i)}_n)$:
 This is the  $i$-th ($i\in \{1,2\}$) evaluation algorithm. It  takes
  the public key $\sf pk$, a labeled program ${\cal P}=(f,\tau_1,\ldots,\tau_n)$, and $n$ ciphertexts $C^{(i)}_1,\dots,C^{(i)}_n$ (labeled by $\tau_1,\ldots,\tau_n$, respectively) as input.
   It  outputs a ciphertext $C^{(i)}$.
	
$\mathsf{2S.Dec(sk},\mathcal{P},C^{(1)},C^{(2)})$: This is a   decryption algorithm. It  takes  the secret key $\sf sk$, a labeled program $\mathcal{P}$, and two ciphertexts $C^{(1)},C^{(2)}$ as input,
 and   outputs a  message $m\in\mathcal{M}$.
\vspace{2mm}

In our model, the client will  run  $\mathsf{2S.KeyGen(1^\lambda)}$ and generate the secret key $\mathsf{sk}$ and
the public key $\mathsf{pk}$. The client runs
 $\mathsf{2S.Enc(sk,\tau},m)$ to encrypt any message $m$ and
  upload the label-encrypted data $C^{(1)}, C^{(2)}$ to the two servers respectively.
  In order to compute a function $f$ on the data items with labels
  $\tau_1,\ldots,\tau_n$, the client simply   sends the program
     $\mathcal{P}=(f,\tau_1,\dots,\tau_{n})$
     to the servers. For every $i\in \{1,2\}$, the $i$th server runs
      $\mathsf{2S.Eval}_i$ to compute a partial result   $C^{(i)}$ for the client. Finally,
      the client runs
      $\mathsf{2S.Dec(sk},\mathcal{P},C^{(1)},C^{(2)})$
      to get the value of  $f(m_1,\dots,m_n)$.

A
\textsf{2S-DCLED} scheme should  satisfy the following properties:    {correctness, succinctness, semantic security} and  {context hiding}.

Informally, the    {correctness} property   requires that whenever the algorithms  $\mathsf{2S.KeyGen}$, $\mathsf{2S.Enc}$,
$\mathsf{2S.Eval_1}$, $\mathsf{2S.Eval_2}$ and $\mathsf{2S.Dec}$ are performed
  correctly, then the client should be able to get the correct value of $f(m_1,\dots,m_n)$.

\begin{Def}{\em (\textbf{Correctness})}
The scheme  $\textsf{2S-DCLED}$ is said to
{\em correctly} evaluate   a function family  $\mathcal{F}$ if for all honestly generated keys
$\mathsf{(sk,pk){\leftarrow}2S.KeyGen(1^\lambda)}$,   for all function $f\in \mathcal{F}$, for all labels
$\tau_{1},\dots,\tau_n \in  \{0,1\}^*$,   for all  messages
  $m_1,\dots,m_n\in\mathcal{M}$, for all ciphertexts
$(C^{(1)}_i,C^{(2)}_i)\leftarrow\mathsf{2S.Enc}(\mathsf{sk},\tau_{i},m_i)$ (where $i=1,2,\ldots,n$),
we have that
$\mathsf{2S.Dec(sk,\mathcal{P},2S.Eval_1(pk,}\mathcal{P},C^{(1)}_1,\dots,C^{(1)}_n),\mathsf{2S.Eval_2(pk,}
\mathcal{P}, \\
	C^{(2)}_1,\dots,C^{(2)}_n))=f(m_1,\dots,m_n).$
\end{Def}

Informally, the   {succinctness} property requires that the size of every ciphertext   should be bounded by some fixed polynomial in the
 security parameter, which is independent of the size of the
 function.
\begin{Def}{\em (\textbf{Succinctness})}
	The $\textsf{2S-DCLED}$ is said to {\em succinctly} evaluate a function family  $\mathcal{F}$ if there is
 a fixed polynomial  $p( \cdot)$ such that every honestly generated ciphertext
  (output of either ${\sf Enc}$ or $\mathsf{Eval}_i$) has size (in bits) $p(\lambda)$.
\end{Def}
The two-server delegation of computation on encrypted data scheme\cite{CF14} (\textsf{2S-DCED}, for short)
  is said to {\em compactly} evaluate
  $\cal F$ if the running time of   decryption is bounded by a fixed polynomial in
  $\lambda$, which is independent of $f$. Although our  {succinctness} property is
  weaker than the  {compactness} property of \cite{CF14}, it is especially
  meaningful when the client is short of communication bandwidth.

Informally, the  {semantic security}    requires that as long as the two servers do not collude
with each other, each individual server cannot  learn any information about
the encrypted data items.
\begin{Def}{\em(\textbf{Semantic Security})}
	  The semantic security of \textsf{2S-DCLED} is defined with the following security game
$\mathbf{Exp}^{\mathsf{SS}}_{\mathsf{2S-DCLED},\cal A}(\lambda)$ between
 a challenger and the PPT adversary $\mathcal{A}$, where
  $\mathcal{A}$ is either the first server or the second server.

		 \textsf{Setup}. The challenger runs $\mathsf{2S.KeyGen(1^\lambda)}$ to obtain a pair
 $\mathsf{(sk,pk)}$ of secret key and public key. It gives the public key
  $\mathsf{pk}$ to  $\cal A$, and keeps the secret key
  $\mathsf{sk}$. It also initializes a list
  $T=\emptyset$ for tracking the queries from $\cal A$.

		 \textsf{Queries}. The adversary $\cal A$ adaptively
issues encryption queries to the challenger, each of the form $(\tau,m)$
 where $\tau\in\{0,1\}^*$ and $m\in\mathcal{M}$. The challenger then proceeds as follows: If $\tau\notin T$, the challenger computes
$(C^{(1)},C^{(2)})\leftarrow\mathsf{2S.Enc}(\mathsf{sk},\tau,m)$,
 updates the list $T=T\cup\{\tau\}$. If  $\mathcal{A}$ is
 the first server, the challenger gives $C^{(1)}$ to $\mathcal{A}$,
 otherwise it gives $C^{(2)}$ to $\mathcal{A}$. If $\tau\in T$, the challenger rejects the query.
	
\textsf{Challenge}. The adversary $\mathcal{A}$ submits a label $\widehat{\tau}\in\{0,1\}^*$ and two data items $m_0$, $m_1$ $\in\cal M$, where $\widehat{\tau}$ is not already in the list $T$. The challenger selects a random bit $B\in\{0,1\}$, computes $(\widehat{C^{(1)}},\widehat{C^{(2)}})\leftarrow\mathsf{2S.Enc}(\mathsf{sk},\widehat{\tau},m_B)$. Same as before, if $\mathcal{A}$ is the first server, the challenger gives $\widehat{C^{(1)}}$ to $\mathcal{A}$, otherwise it gives $\widehat{C^{(2)}}$ to $\mathcal{A}$.
		
\textsf{Output}. The adversary $\mathcal{A}$ outputs $B'$ representing its guess for $B$. $\mathcal{A}$ wins the game if $B'=B$.
		
The advantage $\mathbf{Adv}^{\mathsf{SS}}_{\mathsf{2S-DCLED},\cal A}(\lambda)$ of the adversary $\cal A$ in  this game is defined as $\left|\Pr[B'=B]-\frac{1}{2}\right|,$ where the probability is taken over the random bits used by the challenger and the adversary $\cal A$. We say that the $\textsf{2S-DCLED}$ is semantically secure if for any PPT adversary $\cal A$ it holds $\mathbf{Adv}^{\mathsf{SS}}_{\mathsf{2S-DCLED},\cal A}(\lambda)=\mathsf{negl}(\lambda)$ .

\end{Def}

The {context hiding} property requires  that a receiver computing
 $m\leftarrow \mathsf{2S.Dec(sk},\mathcal{P},C^{(1)},C^{(2)})$ should not be able to learn any additional
 information about  the data $m_1,\dots,m_n$, except what implied by
  $m=f(m_1,\dots,m_n)$.
  Our  {context hiding} property will be defined in a computational setting and different from
  that of  \cite{CF14}.  It is meaningful as the receiver is computationally bounded.
\begin{Def}{\em(\textbf{Context Hiding})}
	{We say that a} \textsf{2S-DCLED} {scheme satisfies {\em context hiding} for function
 family $\mathcal{F}$ if there exists a PPT simulator}
 \textsf{Sim}    {such that the following holds.}
  {For any} $\lambda\in\mathbb{N}$, { any keys}
   $\mathsf{(sk,pk){\leftarrow}2S.KeyGen(1^\lambda)}$,
    {any function} $f\in \mathcal{F}$ {with n inputs},
     {any    messages} $m_1,\dots,m_n\in\mathcal{M}$,
     {any labels} $\tau_1,\dots,\tau_n\in\{0,1\}^*$, {if}
      $(C^{(1)}_i,C^{(2)}_i)\leftarrow\mathsf{2S.Enc(sk},{\tau_i},{m_i})$
       for every $i\in[n]$ and $C^{(i)}=\mathsf{2S.Eval}_i(\mathsf{pk},{\cal P},C^{(i)}_1,\dots,C^{(i)}_n) $
        for $i=1,2$, then  $ \mathsf{Sim}(1^\lambda,\mathsf{sk},\mathcal{P},m)$ is computationally
        indistinguishable from  $(C^{(1)},C^{(2)})$.
\end{Def}

\subsection{Two-Server Verifiable Delegation of Computation on Label-Encrypted Data}

Our definition of
\textsf{2S-DCLED}  has a verifiable  version called
two-server verifiable delegation of computation on label-encrypted data
(\textsf{2S-VDCLED}), which additionally allows the client to verify the
servers' results before actually doing the decryption.
Such a scheme is defined and constructed such that
no dishonest server should be able to persuade the client to accept and output
a wrong value for the outsourced computation.
Formally, a two-server verifiable delegation of computation on label-encrypted data  scheme \textsf{2S-VDCLED} consists of  the following algorithms:

\vspace{2mm}
$ \mathsf{2V.KeyGen(1^\lambda)}$: This is a  key generation algorithm.
 It takes   the security parameter $\lambda$ as input and produces  a secret key $\mathsf{sk}$ and a public key $\mathsf{pk}$ .
	
$ \mathsf{2V.Enc(sk,\tau},m)$: This is an   encryption algorithm.
 It takes the secret key  $\mathsf{sk}$,   any message $m\in {\cal M}$ and its label
 $\tau$ as input, and  outputs two ciphertexts $C^{(1)}$ and $C^{(2)}$.
	
$ \mathsf{2V.Eval}_i(\mathsf{pk},\mathcal{P},C^{(i)}_1,\dots,C^{(i)}_n)$:
 This is the  $i$-th ($i\in \{1,2\}$) evaluation algorithm. It  takes
  the public key $\sf pk$, a labeled program ${\cal P}=(f,\tau_1,\ldots,\tau_n)$, and $n$ ciphertexts $C^{(i)}_1,\dots,C^{(i)}_n$ (labeled by $\tau_1,\ldots,\tau_n$, respectively) as input.
   It  outputs a ciphertext $C^{(i)}$.
	
$\mathsf{2V.Dec(sk},\mathcal{P},C^{(1)},C^{(2)})$: This is a  decryption algorithm.
It  takes  the secret key $\sf sk$,
 a labeled program $\mathcal{P}$, and two ciphertexts
  $C^{(1)},C^{(2)}$ as input, and
  verifies the correctness of  $C^{(1)},C^{(2)}$.
   If both ciphertexts are correct, it decrypts and  outputs a value $m\in\mathcal{M}$.
   Otherwise,  it outputs   $\perp$ to show decryption failure.
\vspace{2mm}

We require   \textsf{2S-VDCLED}  to
 satisfy the properties of
{correctness, succinctness, semantic security}  and {context hiding}.
The definitions of these properties for \textsf{2S-VDCLED} are   similar to those for \textsf{2S-DCLED} and  omitted from here.
An additional property that should be satisfied by {\sf 2S-VDCLED} is
 {unforgeability}, which informally
    requires   that no malicious server should be able to provide wrong responses and
persuade  the client to output a wrong value.
\begin{Def}{\em(\textbf{Unforgeability})}
\label{def:unf}
	The {\em unforgeability} of the  {\sf 2S-VDCLED} scheme  is defined with  the
following security game $\mathbf{Exp}^{\mathsf{UF}}_{\mathsf{2S-VDCLED},\mathcal{A}}(\lambda)$ between a challenger and a  PPT adversary $\mathcal{A}$,
which either plays the role of a malicious first server or a malicious second server:

\textsf{Setup}. The challenger runs $\mathsf{2V.KeyGen(1^\lambda)}$ to obtain
 a pair $\mathsf{(sk,pk)}$ of secret key and public key.
  It gives the public key $\mathsf{pk}$ to   $\mathcal{A}$,
  and keeps the secret key $\mathsf{sk}$. It also
   initializes a list $T=\emptyset$ for tracking the queries from $\mathcal{A}$.

\textsf{Ciphertext Queries}. The adversary $\mathcal{A}$
adaptively queries for the ciphertexts on the pairs of label and message
 of its choice. Given a query $(\tau,m)$ where $\tau\in\{0,1\}^*$ and
  $m\in\mathcal{M}$, the challenger performs the following: If $\tau\notin T$, the challenger computes $(C^{(1)},C^{(2)})\leftarrow\mathsf{2V.Enc}(\mathsf{sk},\tau,m)$,
  updates the list $T=T\cup\{\tau\}$. If $\mathcal{A}$ plays the role of a malicious   first server, the
		challenger gives $C^{(1)}$ to $\mathcal{A}$, otherwise it gives $C^{(2)}$ to $\mathcal{A}$.

\textsf{Verification queries}. The adversary $\mathcal{A}$ adaptively issues verification queries.
Let $(\mathcal{P},C^{(1)})$ or $(\mathcal{P},C^{(2)})$ be a query from $\cal A$,
 where   $\mathcal{P}=({f},{\tau_1},\dots,{\tau_n})$.
 Based on the types of the query, the challenger proceeds as follows.

 Type 1: There exists an index  $i\in[n]$ such that $\tau_i\notin T$, i.e., at least one label
 has not been queried.  If  $\mathcal{A}$ plays the role of a malicious  first server and queries
 with $(\mathcal{P},C^{(1)})$, the challenger sets
  $C^{(2)}=0$ and responds  with the output
 of $\mathsf{2V.Dec(sk},\mathcal{P},{C^{(1)}},{C^{(2)}})$.
  If   $\mathcal{A}$ plays the role of a malicious   second server and queries with $(\mathcal{P},C^{(2)})$,
   the challenger sets $C^{(1)}=0$ and responds with the output of $\mathsf{2V.Dec(sk},\mathcal{P},{C^{(1)}},{C^{(2)}})$.

 Type 2:  $T$ contains all the labels ${\tau_1},\dots,{\tau_n}$. If   $\mathcal{A}$ plays the role of a malicious first server and queries with $(\mathcal{P},C^{(1)})$, the challenger executes $\mathsf{2V.Enc}$ and $\mathsf{2V.Eval}_2$ to compute the $C^{(2)}$ and responds with the output of $\mathsf{2V.Dec(sk},\mathcal{P},{C^{(1)}},{C^{(2)}})$. If $\mathcal{A}$ plays the role of a malicious second server, the challenger executes $\mathsf{2V.Enc}$     and $\mathsf{2V.Eval}_1$ to compute the $C^{(1)}$ and responds with the output of $\mathsf{2V.Dec(sk},\mathcal{P},{C^{(1)}},{C^{(2)}})$.

\textsf{Output}:   $\mathcal{A}$ outputs a forgery ciphertext $\widehat{C^{(1)}}$ or $\widehat{C^{(2)}}$ and a labeled program $\mathcal{P}=(\wh{f},\wh{\tau_1}, \dots,\wh{\tau_n})$. The challenger runs the algorithm  $ \mathsf{2V.Dec(sk},\mathcal{P},C^{(1)},C^{(2)})$ to produce an output $\hat{m}$. $\cal A$ wins the game if $\hat{m}\neq \perp$ and any of the following holds:

Type 1 forgery: There exists an index $i\in[n]$ such that $\wh{\tau_i}\notin T$, i.e., at least one
label $\widehat{\tau}_i$ has not been queried in the game.

Type 2 forgery:  $T$ contains all of the labels $\wh{\tau_1},\dots,\wh{\tau_n}$
 for  the data items $m_1,\dots,m_n$, and $\wh{m}\neq \wh{f}(m_1,\dots,m_n)$, i.e., $\wh{m}$ is not the correct output of program $\mathcal{P}$ when executed on $(m_1,\dots,m_n)$.

The advantage $\mathbf{Adv}^{\mathsf{UF}}_{\mathsf{2S-VDCLED},\mathcal{A}}(\lambda)$ of $\cal A$ in this game is defined as the probability that $\mathcal{A}$ wins. The scheme is said to be {\em existentially unforgeable} under adaptive chosen message and query verification attack, if for all PPT adversaries $\mathcal{A}$,   $\mathbf{Adv}^{\mathsf{UF}}_{\mathsf{2S-VDCLED},\mathcal{A}}(\lambda)=\mathsf{negl}(\lambda)$.

\end{Def}
		\begin{remark}
	In security game of Definition \ref{def:unf}, the adversary $\cal A$ can pose a
verification query of the form  $(\mathcal{P}=(\wh{f},\wh{\tau_1},\dots,\wh{\tau_n}),\wh{C^{(1)}})$
 or $(\mathcal{P}=(\wh{f},\wh{\tau_1},\dots,\wh{\tau_n}),\wh{C^{(2)}})$. $\cal A$ can also terminate
  the \textsf{Verification queries} phase if the response by the challenger is not $\perp$ and any
  of the two types of forgeries happens.
\end{remark}
\begin{remark}
In our treatment of type $i$($i=1,2$) queries, if the adversary ${\cal A}$ plays the role of server i, then the ciphertext $C^{(3-i)}$ will be set to 0 and used for executing $\mathsf{2V.Dec}(\mathsf{sk},{\cal P},C^{(1)},C^{(2)})$.
In fact, it is not always possible to extract the message m encrypted in $C^{(i)}$, as the $C^{(i)}$
is chosen by ${\cal A}$ in a malicious way and  possibly not well-formed.
 Our \textsf{\em 2S-VDCLED} schemes will verify each server's response separately.
  As a result, it does not matter which $C^{(3-i)}$ will be used in decryption.
By default, we set  $C^{(3-i)}=0$.
\end{remark}

\section{A Construction of 2S-DCLED}
In this section we present a construction of two server delegation of computation on label-encrypted data scheme that supports the evaluation of quadratic polynomials on outsourced data. In this scheme, the message space $\mathcal{M}$ is $\mathbb{Z}_p$, where $p$ is a $\lambda$-bit prime. Without loss of generality, we suppose that

\begin{align}
f(x_1,\dots,x_n)=\sum_{i,j\in[n]}\alpha_{i,j}x_ix_j+\sum_{k\in[n]}\beta_{k}x_k+\gamma
\end{align}
is the quadratic polynomial that will be computed in our scheme, where
$\{\alpha_{i,j}\}_{i,j\in[n]}, \{\beta_k\}_{k\in[n]}, \gamma$ are all
 coefficients of $f$ and belong to $\mathbb{Z}_p$. We encrypt
  a message $m_i$ via a PRF $F: {\cal K}\times \{0,1\}^*\rightarrow \mathbb{Z}_p$ and a label $\tau_i$ as $m_i-a_i$ or
  $m_i-b_i$,  where $a_i=F_K(\tau_i\|0)$, $b_i=F_K(\tau_i\|1)$.
   Since we use the label of each message to encrypt the message,
   we call the resulting ciphertext {\em label}-encrypted data.

The computation of any quadratic term $m_1m_2$ will be based on the following mathematical formula
\begin{equation*}
\begin{split}
m_1m_2=&(m_1-a_1)(m_2-a_2)-(a_1-b_1)(a_2-b_2)\\
&+a_1(m_2-b_2)+a_2(m_1-b_1)+b_1b_2.
\end{split}
\end{equation*}
In our scheme, the client will send $(m_1-a_1, a_1-b_1)$, $(m_2-a_2, a_2-b_2)$ to the
first server and ask the first server to compute $C^{(1)}=(m_1-a_1)(m_2-a_2)-(a_1-b_1)(a_2-b_2)$.
 The client will send  $(m_1-b_1,a_1)$ , $(m_2-b_2,a_2)$ to the second server and ask the
 second server to compute  $C^{(2)}=a_1(m_2-b_2)+a_2(m_1-b_1)$. Finally,
  the client can  simply compute  $C^{(1)}+C^{(2)}+b_1b_2$ to learn
  $m_1m_2$. In our scheme, the numbers $a_1$, $a_2$, $b_1$, $b_2$ will
  be pseudorandom values generated with a PRF. As a result, the data on each server will be pseudorandom
     and
    our scheme will be  semantically secure under Definition  4.
In order to compute $m_1+m_2$, the client simply asks  the second server to return
$C^{(2)}=m_1-b_1+m_2-b_2$  and outputs $C^{(1)}+C^{(2)}+(b_1+b_2)$, where $C^{(1)}=0$
as the first server is  idle.

As demonstrated above, in our   \textsf{2S-DCLED} scheme
 the first server will be    responsible to compute the quadratic terms of $f(m_1,m_2,\ldots,m_n)$,
  the     second server will be responsible to compute the linear terms
   of $f(m_1,m_2,\ldots,m_n)$, and finally  the client will be able to  extract the value of
     $f(m_1,m_2,\ldots,m_n)$
     by computing $C^{(1)}+C^{(2)}+f(b_1,\dots,b_n)$. Throughout the process,
        the client only learns  some random values $\{b_i\}_{i=1}^n$
        and the output $f(m_1,\dots,m_n)$, but no
               information about $\{m_i\}_{i=1}^n$.
Our {\sf 2S-DCLED} scheme can be detailed as follows:\vspace{2mm}

 $\mathsf{2S.KeyGen(1^\lambda)}$: Let $p$ be a $\lambda$-bit prime number. Choose a random seed $K\leftarrow{\mathcal{K}}$ for the PRF $F:{\mathcal{K}}\times\{0,1\}^*\rightarrow \mathcal{M}$. Output the secret key $\mathsf{sk}$=$K$ and the public key $\mathsf{pk}$=$p$. The $\mathsf{pk}$ implicitly defines the message space ${\cal M}=\mathbb{Z}_p$.

$\mathsf{2S.Enc(sk},\tau,m)$: Given the secret key $\mathsf{sk}=K$, the  message $m\in\mathcal{M}$ and the label  $\tau\in\{0,1\}^*$, compute $a=F_K(\tau\|0)$ and  $b=F_K(\tau\|1)$, and output $C^{(1)}=(m-a,a-b)$ and $C^{(2)}=(m-b,a)$.

$\mathsf{2S.Eval_1(pk,\mathcal{P},}C^{(1)}_1,\dots,C^{(1)}_n)$: Given the public key $\mathsf{pk}$, a labeled program $\mathcal{P}=(f,\tau_1,\dots,\tau_n)$ and
the ciphertexts $C^{(1)}_1,\dots,C^{(1)}_n$ with labels $\tau_1, \ldots, \tau_n$, output
 $$C^{(1)}=\sum_{i,j\in[n]}\alpha_{i,j}[(m_i-a_i) (m_j-a_j)-(a_i-b_i) (a_j-b_j)].$$	

 $\mathsf{2S.Eval_2}(\mathsf{pk},\mathcal{P},C^{(2)}_1,\dots,C^{(2)}_n)$:
Given the public key $\mathsf{pk}$, a labeled program and the ciphertexts  $C^{(2)}_1,\dots,C^{(2)}_n$ with labels $\tau_1, \ldots, \tau_n$, output
\begin{equation*}	
\begin{split}
	C^{(2)}=&\sum_{i,j\in[n]}\alpha_{i,j}[a_j (m_i-b_i)+a_i(m_j-b_j)]+\\
	&\hspace{0.8mm}\sum_{k\in[n]}\beta_{k}(m_k-b_k).
	\end{split}
\end{equation*}

 $\mathsf{2S.Dec(sk},\mathcal{P},C^{(1)},C^{(2)})$: Given the secret key $\mathsf{sk}$,
a labeled program $\mathcal{P}=(f, \tau_1,\dots,\tau_n)$, and the ciphertexts
$C^{(1)}$ and $C^{(2)}$, compute $b_i\leftarrow F_K(\tau_i\|1)$ for every $i\in [n]$, compute
 $b=f(b_1,\dots,b_n)$, and output
	$m=C^{(1)}+C^{(2)}+b.$
\vspace{2mm}

\noindent
{\bf Correctness.}
The correctness of \textsf{2S-DCLED}  requires that
the algorithm $\sf 2S.Dec$ always outputs the correct value of
the delegated computation, if the scheme is faithfully executed.
\begin{Theo}
	The proposed \textsf{\em 2S-DCLED} scheme is correct.
\end{Theo}

\begin{proof}
 Let $m=C^{(1)}+C^{(2)}+b$ be the output of $\sf 2S.Dec$. If the scheme was
 faithfully executed, then we have
 	\begin{align*}
	m	=&\sum_{i,j\in[n]}\alpha_{i,j}[(m_i-a_i) (m_j-a_j)-(a_i-b_i) (a_j-b_j)]+\\
	&\sum_{i,j\in[n]}\alpha_{i,j}[a_j (m_i-b_i)+a_i (m_j-b_j)]+\\
	&\hspace{0.8mm}\sum_{k\in[n]}\beta_{k}(m_k-b_k)+f(b_1,\dots,b_n)\\
	=&\sum_{i,j\in[n]}\alpha_{i,j}[(m_i-a_i) (m_j-a_j)-(a_i-b_i) (a_j-b_j)+\\
	&\hspace{10mm}a_j (m_i-b_i)+a_i (m_j-b_j)]+\\
&\hspace{1mm}\sum_{k\in[n]}\beta_{k}(m_k-b_k)+f(b_1,\dots,b_n)\\
	=&\sum_{i,j\in[n]}\alpha_{i,j}m_i  m_j+\sum_{k\in[n]}\beta_km_k+\gamma-\sum_{i,j\in[n]}\alpha_{i,j}b_i  b_j-\\
	&\hspace{0.8mm}\sum_{k\in[n]}\beta_kb_k-\gamma +f(b_1,\dots,b_n)\\
	=&f(m_1,\dots,m_n)
	\end{align*}
By Definition 1, our scheme is correct.
\end{proof}

\noindent
\textbf{Semantic Security.}  The semantic security requires that
 each server  learns no information  about the encrypted messages,
 as long as the two servers do not collude with each other.
\begin{Theo}
	If $F$ is a secure  PRF, then the proposed \textsf{\em 2S-DCLED} scheme is semantically secure.
\end{Theo}

\begin{proof}
	We prove the theorem with two games {\bf Game 0} and {\bf Game 1}. Let $W_0$ and $W_1$ be the events that a PPT adversary $\cal A$ wins the semantic security game in {\bf Game 0} and {\bf Game 1}, respectively.

\vspace{1mm}
\noindent	
\textbf{Game 0}: This is   the security game $\mathbf{Exp}^{\mathsf{SS}}_{\mathsf{2S-DCLED},\cal A}(\lambda)$
and defined in  Definition 3.

\vspace{1mm}
\noindent	
	\textbf{Game 1}: This is the same as {\bf Game 0}, except that the PRF $F$ is replaced by a truly { random function}. That is, the challenger chooses  $a,b\leftarrow \cal M$ instead of computing $a=F_K(\tau\|0)$ and $b=F_K(\tau\|1)$ in the \textsf{2S.Enc} procedure. It's easy to see that there is a PRF adversary $\mathcal{B}$ such that:
	\begin{align}
	\left|\Pr[W_0]-\Pr[W_1]\right|\leq \mathbf{Adv}^{PRF}_{F,\mathcal{B}}(\lambda),
	\end{align}
where   $\mathbf{Adv}^{PRF}_{F,\mathcal{B}}(\lambda)$ is  the advantage of  $\mathcal{B}$ winning the PRF security game.

	If $\cal A$ is the first server, $\wh{C^{(1)}}=(m_B-\wh{a},\wh{a}-\wh{b})$.
Since $\wh{a}$ and $\wh{b}$ are random values in $\cal M$, $m_B-\wh{a}$ and
$\wh{a}-\wh{b}$ are independently and uniformly distributed over $\cal M$.
If $\cal A$ is the second server, $\wh{C^{(2)}}=(m_B-\wh{b},\wh{a})$.
Since $\wh{a}$ and $\wh{b}$ are random values in $\cal M$, $m_B-\wh{b}$
 and $\wh{a}$ are also independently and uniformly distributed over $\cal M$.
  Hence, in both cases we have that
	\begin{align}
	\Pr[W_1]=\frac{1}{2}
	\end{align}
	Putting together equations (2), (3) and (4), we will have that
	\begin{align*}
	\mathbf{Adv}^{\mathsf{SS}}_{\mathsf{2S-DCLED},\cal A}(\lambda)=\left|\Pr[W_0]-\frac{1}{2}\right|\leq \mathbf{Adv}^{PRF}_{F,\mathcal{B}}(\lambda),
	\end{align*}
which  completes the proof.
\end{proof}

\noindent
\textbf{Context Hiding}.
The context hiding property requires that the
 receiver running $\mathsf{2S.Dec(sk},\mathcal{P},C^{(1)},C^{(2)})$
 should learn  no additional information about the data $m_1,\dots,m_n$,
 except   what implied by $f(m_1,\dots,m_n)$.
\begin{Theo}
	The proposed  \textsf{\em 2S-DCLED} scheme  satisfies the context hiding property.
\end{Theo}
\begin{proof}
By Definition 5, we need to construct a  simulator $\mathsf{Sim}(1^\lambda,\mathsf{sk},\mathcal{P},m)$ that takes the secret
key $\sf sk$, the program $\cal P$ and the scheme's output $m$ as input such that its output
is a pair   $(\widehat{C^{(1)}},\widehat{C^{(2)}})$ of ciphertexts that is
computationally  indistinguishable from the servers' responses
   $({C^{(1)}},{C^{(2)}})$
  in a real execution of the proposed scheme.

When  $\deg(f)=1$, we have that
$\alpha_{i,j}=0$ for all $i,j\in[n]$.
It's easy  to see that
 $C^{(1)}=0$ and
$C^{(2)}=\sum_{k\in[n]}\beta_{k}(m_k-b_k)$  is pseudorandom
 over ${\cal M}=\mathbb{Z}_p$, where $b_k=F_K(\tau_k\|1)$
for every $k\in[n]$. As $m=C^{(2)}+f(b_1,b_2,\ldots,b_n)$, our simulator
$\sf Sim$ will output
$\wh{C^{(1)}}=0$ and $\wh{C^{(2)}}=m-f(b_1,b_2,\ldots,b_n)$.
It's easy to see that   ${\sf Sim}(1^\lambda,\mathsf{sk},\mathcal{P},m)$
and $(C^{(1)},C^{(2)})$ are identically distributed, which implies that both distributions are computationally indistinguishable.
  When $\deg(f)=2$, it is not hard to see that
  $C^{(i)}$  is pseudorandom over  ${\cal M}$ for every
  $i\in\{1,2\}$
  and $C^{(1)}+C^{(2)}=m-f(b_1,b_2,\ldots,b_n)$.
  Our simulator $\sf Sim$ will choose $\wh{C^{(2)}}\leftarrow {\cal M}$
  uniformly at random, compute
  $\wh{C^{(1)}}=m-f(b_1,b_2,\ldots,b_n)-\wh{C^{(2)}}$,
  and output  $(\wh{C^{(1)}}, \wh{C^{(2)}})$.
It's easy to see that
$(\wh{C^{(1)}}, \wh{C^{(2)}})$ and $(C^{(1)}, C^{(2)})$
are computationally indistinguishable.
\end{proof}

\section{A Construction of 2S-VDCLED}
In this section, we present a construction of two server verifiable delegation of computation on label-encrypted data scheme (\textsf{2S-VDCLED}) that supports the verifiable evaluation of quadratic polynomials. In this construction the message space $\cal M$ is $\mathbb{Z}_p$, where $p$ is a $\lambda$-bit prime. We define $f:{\cal M}^n \rightarrow \cal M$ as equation (1).

We use the homomorphic MACs \cite{CF18} to achieve verification. In the homomorphic MACs of \cite{CF18}, the authentication tag of a message $m\in\mathcal{M}$ with label $\tau\in\{0,1\}^*$ is a linear polynomial $y(x)\in\mathbb{Z}_p[x]$  such that $y(0)=m$ and $y(s)=r_{\tau}$, where $r_{\tau}=F_K(\tau)$.
These operations are naturally
homomorphic with respect to the evaluation of the polynomial at every point.  In particular, if we have two tags $y^{(1)}$ and $y^{(2)}$ such that $y^{(1)}(0)=m_1$ and $y^{(2)}(0)=m_2$, then for $y=y^{(1)}+y^{(2)}$ (resp. $y=y^{(1)}y^{(2)}$) we clearly have $y(0)=m_1+m_2$ (resp. $y(0)=m_1m_2$). The same homomorphic property holds for its evaluation at the random point $s$, i.e., $y(s)=r_{\tau_1}+r_{\tau_2}$ (resp. $y(s)=r_{\tau_1}r_{\tau_2}$). By extending this argument to the evaluation of a function  $f$, this allows to verify a tag $y$ for a labeled program $\mathcal{P} = (f,\tau_1,\dots,\tau_n)$ and a message $m$, by simply checking that $m=y(0)$ and $f(r_{\tau_{1}},\dots,r_{\tau_{n}})=y(s)$, where $r_{\tau_i}=F_K(\tau_{i})$ for all $i\in[n]$.

We use homomorphic MACs separately for each server, to ensure that the output of each server is correct.
Below is the description of our {\sf 2S-VDCLED} scheme.

\vspace{2mm}
$\mathsf{2V.KeyGen(1^\lambda)}$: Let $p$ be a $\lambda$-bit prime. Choose two random seeds
 $(K_1,K_2)\leftarrow\mathcal{K}^2$ for a PRF $F:{\cal K}\times\{0,1\}^*\rightarrow {\cal M}$.
  Choose   $(s_1,s_2)\leftarrow\mathbb{Z}^2_{p}$. Output the secret
  key $\mathsf{sk}=(K_1,K_2,s_1,s_2)$ and  the public key $\mathsf{pk}=p$.
   The $\mathsf{pk}$ implicitly defines   ${\cal M}=\mathbb{Z}_p$.
	
$\mathsf{2V.Enc(sk,\tau},m)$: Given the secret key $\mathsf{sk}=(K_1,K_2,s_1,s_2)$,
proceed as follows to encrypt any message $m \in \mathbb{Z}_p$ with label $\tau\in\{0,1\}^*$.
First, compute
	$a=F_{K_1}(\tau\|0)$, $b=F_{K_1}(\tau\|1)$, $r_1=F_{K_2}(\tau\|0)$, $r_2=F_{K_2}(\tau\|1)$,
$r_3=F_{K_2}(\tau\|2)$ and $r_4=F_{K_2}(\tau\|3)$. The ciphertext of $m$ consists of four polynomials
 $y^{(1)}$, $y^{(2)}$, $y^{(3)}$, $y^{(4)}$, where $y^{(1)}=(m-a)+\frac{r_1-(m-a)}{s_1}x$, $y^{(2)}=(a-b)+\frac{r_2-(a-b)}{s_1}x$, $y^{(3)}=(m-b)+\frac{r_3-(m-b)}{s_2}x$,
$y^{(4)}=a+\frac{r_4-a}{s_2}x$. The polynomial $y^{(1)}$ is constructed  such that $y^{(1)}(0)=m-a$ and $y^{(1)}(s_1)=r_1$.
  The other polynomials $y^{(2)}$, $y^{(3)}$ and $y^{(4)}$ are   constructed with the same idea. This algorithm outputs $C^{(1)}=(y^{(1)},y^{(2)})$ and $C^{(2)}=(y^{(3)},y^{(4)})$.
	
 $\mathsf{2V.Eval_1(pk, \mathcal{P}, }C^{(1)}_1,\dots,C^{(1)}_n)$:
This algorithm takes the public key $\mathsf{pk}$, a labeled program
$\mathcal{P}=(f,\tau_1,\dots,\tau_n)$ and the ciphertexts  $C^{(1)}_1,\dots,C^{(1)}_n$
(labeled by $\tau_1,\ldots, \tau_n$, respectively) as input, where
$C^{(1)}_i=(y^{(i,1)},y^{(i,2)})$ for every $i\in[n]$. It outputs
$$C^{(1)}=\sum_{i,j\in[n]}\alpha_{i,j}[y^{(i,1)}y^{(j,1)}-y^{(i,2)}y^{(j,2)}],$$
which is a quadratic polynomial in $x$ and usually represented with  the
 field elements $y^{(1)}_0$, $y^{(1)}_1$, $y^{(1)}_2$ such that $C^{(1)}=y^{(1)}_0+y^{(1)}_1x+y^{(1)}_2x^2$.
	
$\mathsf{2V.Eval_2(pk, \mathcal{P}, }C^{(2)}_1,\dots,C^{(2)}_n)$:
This algorithm takes the public key $\mathsf{pk}$, a labeled program
 $\mathcal{P}=(f,\tau_1,\dots,\tau_n)$ and the ciphertexts  $C^{(2)}_1,\dots,C^{(2)}_n$ (labeled by $\tau_1,\ldots, \tau_n$, respectively)
 as input, where   $C^{(2)}_i=(y^{(i,3)},y^{(i,4)})$ for every $i\in[n]$.
 It outputs
	\begin{align*}
	C^{(2)}=\sum_{i,j\in[n]}\alpha_{i,j}[y^{(i,3)}y^{(j,4)}+y^{(i,4)}y^{(j,3)}]+\hspace{0.8mm}\sum_{k\in[n]}\beta_{k}y^{(k,3)},
	\end{align*}
which is a quadratic polynomial in $x$ and usually represented with the field elements $y^{(2)}_0$, $y^{(2)}_1$, $y^{(2)}_2$ such that $C^{(2)}=y^{(2)}_0+y^{(2)}_1x+y^{(2)}_2x^2$.
	
 $\mathsf{2V.Dec(sk},\mathcal{P},C^{(1)},C^{(2)})$:
This algorithm takes
the secret key
 $\mathsf{sk}=(K_1,K_2,s_1,s_2)$, a labeled program
  $\mathcal{P}=(f,\tau_1,\dots,\tau_n)$,
   $C^{(1)}=(y^{(1)}_0,y^{(1)}_1,y^{(1)}_2)\in \mathbb{Z}_p^3$ and
    $C^{(2)}=(y^{(2)}_0,y^{(2)}_1,y^{(2)}_2)\in \mathbb{Z}_p^3$ as input.
It computes $b_i=F_{K_1}(\tau_i\|1)$, $r_{i,1}=F_{K_2}(\tau_i\|0)$, $r_{i,2}=F_{K_2}(\tau_i\|1)$,
 $r_{i,3}=F_{K_2}(\tau_i\|2)$ and $r_{i,4}=F_{K_2}(\tau_i\|3)$ for every $i\in [n]$, and sets $R_1=\sum_{i,j\in[n]}\alpha_{i,j}[r_{i,1}r_{j,1}-r_{i,2}r_{j,2}]$,
$R_2=\sum_{i,j\in[n]}\alpha_{i,j}[r_{i,3}r_{j,4}+r_{i,4}r_{j,3}]+\sum_{k\in[n]}\beta_{k}r_{k,3}$. The algorithm  checks whether the following equations hold:
$R_1=y^{(1)}_0+y^{(1)}_1s_1+y^{(1)}_2{s_1^2}$; $R_2=y^{(2)}_0+y^{(2)}_1s_2+y^{(2)}_2{s_2^2}.$
	If both equations hold,  this algorithm computes $b=f(b_1,\dots,b_n)$ and outputs
	$m=y^{(1)}_0+y^{(2)}_0+b.$
Otherwise,  it outputs $\perp$.
\vspace{2mm}

The  proofs for the
{correctness, succinctness, semantic security} and  {context hiding}  properties
of the  $\sf 2S$-$\sf VDCLED$  scheme
are quite similar to those for our {\sf 2S-DCLED} scheme and omitted from here.
It remains to show the unforgeability of the proposed {\sf 2S-VDCLED} scheme.

\vspace{1mm}
\noindent
\textbf{Unforgeability.}  This  property requires  that
no  adversary
that plays the role of a malicious first server or the role of a malicious second server
is able to persuade the client to output a wrong  value for the delegated computations.
\begin{Theo}
	 {Suppose that} $F$  {is an  PRF}. Then  {the proposed
\textsf{\em 2S-VDCLED} scheme is unforgeable. In particular, for any PPT adversary} $\mathcal{A}$ that makes
$\leq Q$ verification queries, we have that
	$$\mathbf{Adv}^{\mathsf{UF}}_{\mathsf{2S-VDCLED},\mathcal{A}}(\lambda)
\leq \epsilon_{F}
	+\frac{2(Q+1)}{p-2Q}$$
	 {where} $\epsilon_F$ is an upper bound on the
 {advantage of any PPT adversary winning the PRF security game with respect to $F$}.
\end{Theo}
\begin{proof}
	We define two games {\bf Game 0} and {\bf Game 1} and let $W_0$, $W_1$ be the events that $\mathcal{A}$ wins in {\bf Game 0} and {\bf Game 1}, respectively.

\vspace{1mm}
\noindent	
\textbf{Game 0}: This game  is the standard security game $\mathbf{Exp}^{\mathsf{UF}}_{\mathsf{2S-VDCLED},\mathcal{A}}(\lambda)$ of Definition 5.
We have that
	\begin{align}
	\Pr[W_0]=\mathbf{Adv}^{\mathsf{UF}}_{\mathsf{2S-VDCLED},\mathcal{A}}(\lambda).
	\end{align}

\vspace{1mm}
\noindent
	\textbf{Game 1}:  This game is identical to {\bf Game 0}, except that the PRF $F$ is replaced with a
truly random function. That is, for every label $\tau$, the challenger generates
$a,b,r_1,r_2,r_3,r_4{\leftarrow}\mathbb{Z}_p$ instead of
computing $a=F_{K_1}(\tau\|0)$, $b=F_{K_1}(\tau\|1)$, $r_1=F_{K_2}(\tau\|0)$, $r_2=F_{K_2}(\tau\|1)$,
 $r_3=F_{K_2}(\tau\|2)$ and $r_4=F_{K_2}(\tau\|3)$. It's trivial to see that
	\begin{align}
	|\Pr[W_1]-\Pr[W_0]|\leq  \epsilon_F.
	\end{align}
	Without loss of generality, we suppose that $\cal A$ plays the role of a malicious
 first server. Then the challenger in Game 1 will work  as follows.

 \vspace{2mm}
	\textsf{Ciphertext Queries.} The adversary submits queries $(\tau_i,m_i)$ where $\tau_i$ is the label of message $m_i$. The challenger creates a new list $T$ for tracking the queries from ${\cal A}$ in the game.
		 For the $i$-th query, if $T$ does not contain  $\tau_i$, i.e., the label $\tau_i$ was never queried. The challenger responds as follows: choose $a_i,b_i,r_{i,1},r_{i,2},r_{i,3},r_{i,4}{\leftarrow}\mathbb{Z}_p$; compute $C^{(1)}_i=((m_i-a_i)+\frac{r_{i,1}-(m_i-a_i)}{s_1}x,(a_i-b_i)+\frac{r_{i,2}-(a_i-b_i)}{s_1}x)$; send $C^{(1)}_i$ to $\mathcal{A}$ and update $T=T\cup{(\tau_i,a_i,b_i,r_{i,1},r_{i,2},r_{i,3},r_{i,4})}$. If $(\tau_i,\cdot,\cdot,\cdot,\cdot,\cdot,\cdot)\in T$, i.e. label $\tau_i$ was previous queried, the challenger rejects the query.

\textsf{Verification Queries}. The adversary submits queries $(\mathcal{P}_i,C^{(1)}_i)$ where program $\mathcal{P}_i=(f_i,(\tau_{i,1},\dots,\tau_{i,n_i}))$. The challenger responds to the $i$-th query as follows: If there is a $j\in[n_i]$ such that $(\tau_{i,j},\cdot,\cdot,\cdot,\cdot,\cdot,\cdot)\notin T$, the challenger chooses random values $a_{i,j},b_{i,j},r_{i,j,1},r_{i,j,2},r_{i,j,3},r_{i,j,4}{\leftarrow}\mathbb{Z}_p$. The challenger let $C^{(2)}=0$ and responds with the output of $\mathsf{Dec(sk},\mathcal{P},{C^{(1)}},{C^{(2)}})$.	If $(\tau_{i,j},\cdot,\cdot,\cdot,\cdot,\cdot,\cdot)\in T$ for every $j\in[n_i]$, the challenger retrieves the random values
			$a_{i,j},b_{i,j},r_{i,j,1},r_{i,j,2},r_{i,j,3},r_{i,j,4}$ corresponding to $\tau_{i,j}$ from $T$. The challenger performs  $\mathsf{2V.Eval}_2$ to compute the $C^{(2)}$ and responds with the output of $\mathsf{Dec(sk},\mathcal{P},{C^{(1)}},{C^{(2)}})$.

	Eventually $\cal A$ outputs $(\mathcal{P},\wh{C^{(1)}})$, where $\mathcal{P}=(\wh{f},\wh{\tau_1},\dots,\wh{\tau_{\hat{n}}})$, $\wh{C^{(1)}}=(\wh{y^{(1)}_0},\wh{y^{(1)}_1},\wh{y^{(1)}_2})$. The adversary  $\cal A$ wins the game if any of the two following types of forgeries occurs.

	 Type 1 forgery: If there exists an  $i\in[\wh{n}]$  $(\wh{\tau_{i}},\cdot,\cdot,\cdot,\cdot,\cdot,\cdot)\notin T$, the challenger chooses random values $\wh{a_i},\wh{b_i},\wh{r_{i,1}},\wh{r_{i,2}},\wh{r_{i,3}},\wh{r_{i,4}}{\leftarrow}\mathbb{Z}_p$ and retrieves
the  random values corresponding to the remaining labels from  $T$. Let $\wh{R_1}=\sum\limits_{i,j\in[\hat{n}]}\wh{\alpha_{i,j}}[\wh{r_{i,1}}\wh{r_{j,1}}-\wh{r_{i,2}}\wh{r_{j,2}}]$. The adversary wins the game if:
		\begin{align}
		\wh{R_1}=\wh{y^{(1)}_0}+\wh{y^{(1)}_1}s_1+\wh{y^{(1)}_2}s_1^2.
		\end{align}
Type 2 forgery: If for all $i\in[\wh{n}]$ such that $(\wh{\tau_{i}},\cdot,\cdot,\cdot,\cdot,\cdot,\cdot)\in T$, the challenger retrieves $(\wh{a_i},\wh{b_i},\wh{r_{i,1}},\wh{r_{i,2}},\wh{r_{i,3}},\wh{r_{i,4}})$ from list $T$. Let $\{({\wh{a_i}},{\wh{b}},{\wh{r_{i,1}}},{\wh{r_{i,2}}},{\wh{r_{i,3}}},{\wh{r_{i,4}}},m_i)\}^{\hat{n}}_{i=1}$ be the  random values and data corresponding to the labels $\wh{\tau_1},\dots,\wh{\tau_{\hat{n}}}$. Let $\wh{b}=f({\wh{b_1}},\dots,{\wh{b_{\hat{n}}}})$,  $\wh{R_1}=\sum\limits_{i,j\in[\hat{n}]}\wh{\alpha_{i,j}}[\wh{r_{i,1}}\wh{r_{j,1}}-\wh{r_{i,2}}\wh{r_{j,2}}]$, $\wh{R_2}=\sum_{i,j\in[\hat{n}]}\wh{\alpha_{i,j}}[\wh{r_{i,3}}\wh{r_{j,4}}+\wh{r_{i,4}}\wh{r_{j,3}}]+\sum_{k\in[\hat{n}]}\wh{\beta_{k}}\wh{r_{k,3}}$. Let $\wh{m}=\wh{y^{(1)}_0}+\wh{y^{(2)}_0}+\wh{b}$. The adversary wins the game if:
		\begin{align}
		\wh{R_1}=\wh{y^{(1)}_0}+\wh{y^{(1)}_1}s_1+\wh{y^{(1)}_2}s_1^2
		\end{align}
		and $\wh{m}\neq m=\wh{f}(m_1,\dots,m_{\hat{n}})$.

	We now compute the probability of $\cal A$ winning {\bf Game 1}. Let $B_i$ be the event that
$\cal A$ wins the game after $i$ verification queries. Let $Q$ be the upper bound on the number of verification queries made by $\cal A$. We have:
	$\Pr[W_1]=\Pr[\bigcup^Q_{i=0} B_i]\leq \sum_{i=0}^{Q}\Pr[B_i].$
	Let $V$, $\neg V$ be the events that $\cal A$ outputs a type 1 forgery and a type 2 forgery, respectively.
	
	Event $V$ happens (type 1 forgery): The left hand side of (6) is a random value in $\mathbb{Z}_p$ that is independent of $\cal A$’s view. In addition, since $s_1$ is a secret key, the probability that equation (7) holds is exactly $2/p$. Hence,
	\begin{align}
	\Pr[B_i\cap V]=\frac{2}{p}\Pr[V]
	\end{align}
	
	Event $\neg V$ happens (type 2 forgery): In this case, $\cal A$ uses a program $\mathcal{P}=(\wh{f},\wh{\tau_1},\cdots,\wh{\tau_{\hat{n}}})$ and  all the labels have been posed in the previous ciphertext queries. Event $B_i$ happens if $\wh{m}\neq \wh{f}(m_1,\dots,m_{\hat{n}})$ and equation (8) holds.
	
	Let ${C^{(1)}_j}$ be the ciphertext corresponding to label $\wh{\tau_j}$ in a previous ciphertext query, for all $j\in[\wh{n}]$.
	Define ${C^{(1)}}=({y^{(1)}_0},{y^{(1)}_1},{y^{(1)}_2})\leftarrow \mathsf{2V.Eval_1(pk,}\mathcal{P},({C^{(1)}_1},\dots,{C^{(1)}_{\hat{n}}}))$. Since ${C^{(1)}}$ is a valid ciphertext for $m=\wh{f}(\wh{\tau_1},\cdots,\wh{\tau_{\hat{n}}})$, the following relation holds:
	\begin{align}
	\wh{R_1}&=\sum\limits_{i,j\in[\hat{n}]}\wh{\alpha_{i,j}}[\wh{r_{i,1}}\wh{r_{j,1}}-\wh{r_{i,2}}\wh{r_{j,2}}]\\
	&={y^{(1)}_0}+{y^{(1)}_1}s_1+{y^{(1)}_2}s_1^2.
	\end{align}
	Subtracting (10) from (7), we obtain:
	\begin{equation}
\label{eqn1}
	\begin{split}
	(\wh{y^{(1)}_0}-{y^{(1)}_0})+(\wh{y^{(1)}_1}-{y^{(1)}_1})s_1+ (\wh{y^{(1)}_2}-{y^{(1)}_2})s_1^2 =0.
	\end{split}
	\end{equation}
	Since $\wh{m}\neq f(m_1,\dots,m_{\hat{n}})$, we know that $\wh{C^{(1)}}\neq {C^{(1)}}$  {implying that the left-hand side of (\ref{eqn1}) is a nonzero polynomial in $s_1$.} Hence, in producing a valid forgery, $\cal A$ must guess secret key $s_1$.
	
	As $s_1$ is uniformly distributed over $\mathbb{Z}_p$, we have $\Pr[B_0\bigcap \neg V]=2/p\cdot \Pr[\neg V]$. After the first verification query, since there are $\leq 2$ values of $s_1$ that satisfy equation (11), the number of possible values for $s_1$ becomes $\geq p-2$. Therefore, after $i$ queries, $\cal A$ can exclude at most $2i$ possible values of $s_1$, meaning that the number of possible values for $s_1$ is at least $p-2i$. Thus,
	\begin{align}
	\Pr[B_i\cap \neg V]\leq \frac{2}{p-2i}\cdot \Pr[\neg V].
	\end{align}
	From equations (9) and (13), we obtain:
	$$\Pr[B_i]\leq\frac{2}{p-2i}(\Pr[V]+\Pr[\neg V])\leq\frac{2}{p-2i}.$$
	Finally, we have
	\begin{align}
	\Pr[W_1]\leq\sum_{i=0}^Q\Pr[B_i]\leq \frac{2(Q+1)}{p-2Q}.
	\end{align}
	Putting together equations (4), (5) and (13),
	$$\mathbf{Adv}^{\mathsf{UF}}_{\mathsf{2S-VDCLED},\mathcal{A}}(\lambda) \leq
\epsilon_F
	+\frac{2(Q+1)}{p-2Q}.$$
	Since $p\approx 2^\lambda$and $Q$ is a polynomial of $\lambda$,
 $\frac{2(Q+1)}{p-2Q}=\mathsf{negl}(\lambda)$ thus completing the proof of Theorem 4.
\end{proof}

\section{Generalization to $d$ Servers $(d>2)$}
Our definitions   for {\sf 2S-DCLED} and {\sf 2S-VDCLED} can be generalized to
 the $d$-server case for any integer $d>2$, which give the models for
 $d${\sf S-DCLED}  and $d${\sf S-VDCLED}.
 In this section, we show how to
 delegate any
  degree-$d$ computations using $d$ non-communicating servers.

\begin{table*}[t]
\caption{The algorithms' running time in \textsf{2S-DCED}, \textsf{2S-DCLED} and \textsf{2S-VDCLED}}
	\centering
	\resizebox{\textwidth}{!}{%
		\begin{tabular}{|c|c|c|c|c|c|c|c|c|c|c|}
			\hline
			& \multicolumn{2}{c|}{The Configuration of f} & \multicolumn{2}{c|}{\sf 2S-DCED} & \multicolumn{2}{c|}{\sf 2S-DCLED} & \multicolumn{2}{c|}{\sf 2S-VDCLED} & \begin{tabular}[c]{@{}c@{}}{\sf 2S-DCED} vs.\\ {\sf 2S-DCLED}\end{tabular} & \begin{tabular}[c]{@{}c@{}}{\sf 2S-DCED} vs.\\ {\sf 2S-VDCLED}\end{tabular} \\ \hline
			Data Size & \begin{tabular}[c]{@{}c@{}}The Number of\\ Quadratic terms\end{tabular} & \begin{tabular}[c]{@{}c@{}}The Number of\\ Liner Terms\end{tabular} & $\mathsf{Eval}$(sec) & $\mathsf{Dec}$(sec) & $\mathsf{Eval}$(sec) & $\mathsf{Dec}$(sec) & $\mathsf{Eval}$(sec) & $\mathsf{Dec}$(sec) & \begin{tabular}[c]{@{}c@{}}Sever\\ Speedup\end{tabular} & \begin{tabular}[c]{@{}c@{}}Sever\\ Speedup\end{tabular} \\ \hline
			10 & 55 & 10 & 0.0221 & 0.0667 & 0.00001 & 0.000019 & 0.000037 & 0.000097 & 2210 & 597.2973 \\ \hline
			50 & 1275 & 50 & 0.5055 & 0.0668 & 0.000229 & 0.000133 & 0.00084 & 0.000687 & 2207.424 & 601.7857 \\ \hline
			100 & 5050 & 100 & 1.9495 & 0.0668 & 0.000889 & 0.000365 & 0.00322 & 0.001908 & 2192.913 & 605.4348 \\ \hline
			500 & 125250 & 500 & 47.7712 & 0.0668 & 0.021744 & 0.005844 & 0.080275 & 0.031103 & 2196.983 & 595.0944 \\ \hline
			1000 & 500500 & 1000 & 191.5906 & 0.0667 & 0.0872 & 0.021892 & 0.317646 & 0.109044 & 2197.14 & 603.1576 \\ \hline
		\end{tabular}%

	}

\end{table*}

\subsection{Basic Ideas for Constructing  $d$S-DCLED}

It suffices to demonstrate the idea for computing any
 degree-$d$ monomial  $f(m_1,m_2,\ldots,m_d)=\prod_{i=1}^{d}m_i$ with $d$ non-communicating servers.
In our construction, the client  will use  a PRF $F$ to generate a  pseudorandom number $a_{i,j}\in {\cal M}$  for
  every  $i,j\in[d]$. For every $j\in[d]$, it stores the following data on the
  $j$-th server.
\begin{center}
  \begin{tabular}{|c|}
    \hline
    Server $j$ \\ \hline
    $a_{1,1},\dots,a_{1,j-1},m_1-a_{i,j},a_{1,j+1},\dots,a_{1,d}$ \\ \hline
    $a_{2,1},\dots,a_{2,j-1},m_2-a_{2,j},a_{2,j+1},\dots,a_{2,d}$ \\ \hline
    $\cdots$ \\ \hline
    $a_{d,1},\dots,a_{d,j-1},m_d-a_{d,j},a_{d,j+1},\dots,a_{d,d}$ \\ \hline
  \end{tabular}
  \end{center}
In our construction, the first server will be responsible to compute $S_1=\prod_{i=1}^{d}(m_i-a_{i,1})$ and set $c_1=1$.
The second server will be responsible to  eliminate the
degree-$(d-1)$ terms in $S_1$.
More precisely, the
  second server will compute  $S_2=\sum_{i_1=1}^{d}\frac{\prod_{i=1}^{d}(m_i-a_{i,2})}{(m_{i_1}-a_{i_1,2})}a_{i_1,1}$
  and set {$c_2=a_{i_1,1}$}. The third server will be responsible  to eliminate the  degree-$(d-2)$ terms in
  both  $S_1$ and $S_2$.
More precisely, it will compute  $S_3=\prod_{1\leq i_1<i_2\leq d}\frac{\prod_{i=1}^{d}(m_i-a_{i,3})}{(m_{i_1}-a_{i_1,3})
  (m_{i_2}-a_{i_2,3})}[c_2(-a_{i_2,1})+c_2a_{i_2,2}]$ and set $c_3=c_2(-a_{i_2,1})+c_2a_{i_2,2}$.
In general, for every $j>1$, the $j$-th server will be responsible to
  eliminate all degree-$(d-j+1)$ terms that arise from   the   computations of
  $S_1,\dots,S_{j-1}$. More precisely,
  it will  compute  $S_j=\sum_{1\leq i_1<\dots<i_{j-1}\leq d}\frac{\prod_{i=1}^{d}
   (m_i-a_{i,j})}{(m_{i_1}-a_{i_1,j})\dots(m_{i_{j-1}}-a_{i_{j-1},j})}[ c_2\prod_{k=2}^{j-1}
   (-a_{i_k,1})\\ +c_2a_{i_2,2}\prod_{k=3}^{j-1}(-a_{i_k,2})+c_3a_{i_3,3}\prod_{k=4}^{j-1}(-a_{i_k,3})+
   \dots+c_{j-1}a_{i_{j-1},j-1} ]$.
   The following theorem shows that based on the servers' responses a client can
   reconstruct $f(m_1,m_2,\ldots,m_d)$ with limited local computations.
\begin{Theo}
	Let $S_j$  be defined as above for every $j\in[d]$.
Let $\mathbb{P}_d=\{{\bm i}=(i_1,\ldots,i_d): \{i_1,\ldots,i_d\}=[d]\}$ be the set of all permutations of
$[d]$. Then
\begin{equation}\sum_{j=1}^{d}S_j=\prod_{i=1}^{d}m_i+\prod_{i=1}^{d}(-a_{i,1})+
\sum_{j=2}^{d}  \sum_{{\bm i}\in \mathbb{P}_d}c_j\prod_{k=j}^{d}(-a_{i_k,j}).
\end{equation}
\end{Theo}
\begin{proof}
We show that   $S_j$ can eliminate   all
degree-$(d-j+1)$ terms in $S_1,\dots,S_{j-1}$ for every $j\in \{2,\ldots,d\}$.
In $S_1$, the  degree-$(d-j+1)$ terms  coefficients
 $c'_{1}=\prod_{k=1}^{j-1}(-a_{i_k,1})=-a_{i_1,1}\prod_{k=2}^{j-1}(-a_{i_k,1})=-c_2\prod_{k=2}^{j-1}(-a_{i_k,1})$.
  In $S_\ell$, for $\ell=2,\dots, j-1$,  the degree-$(d-j+1)$ terms coefficients $c'_\ell=c_\ell\prod_{k=\ell}^{j-1}(-a_{i_k,l})=c_\ell[-a_{i_\ell,l}\prod_{k=\ell+1}^{j-1}(-a_{i_k,l})]$.
	We have known the  degree-$(d-j+1)$ terms coefficients of $S_j$ is $c_j=c_2\prod_{k=2}^{j-1}(-a_{i_k,1})+c_2a_{i_2,2}\prod_{k=3}^{j-1}(-a_{i_k,2})+c_3a_{i_3,3}\prod_{k=4}^{j-1}(-a_{i_k,3})+\dots+c_{j-1}a_{i_{j-1},j-1}$. To prove $S_j$ can eliminate   all
	degree-$(d-j+1)$ terms in $S_1,\dots,S_{j-1}$ for every $j\geq 2$, we just prove $\sum_{\ell=1}^{j-1}c'_\ell+c_j=0$. In fact,
	\begin{align*}
	\sum_{\ell=1}^{j-1}c'_\ell+c_j
	=&-c_2\prod_{k=2}^{j-1}(-a_{i_k,1})+\sum_{\ell=2}^{j-1}c_\ell[-a_{i_\ell,l}\prod_{k=\ell+1}^{j-1}(-a_{i_k,l})]\\
	&+c_2\prod_{k=2}^{j-1}(-a_{i_k,1})+c_2a_{i_2,2}\prod_{k=3}^{j-1}(-a_{i_k,2})\\
&+c_3a_{i_3,3}\prod_{k=4}^{j-1}(-a_{i_k,3})+\dots+c_{j-1}a_{i_{j-1},j-1}\\
	=&0.
	\end{align*}

	Thus,  $\sum_{j=1}^{d}S_j$ only contains the degree-$d$   term $\prod_{i=1}^{d}m_i$  and the constant terms. Next, we give a concrete expression of the constant terms. In $S_1$, the constant term is $\prod_{i=1}^{d}(-a_{i,1})$. For $S_j$, $j=2,\dots, d$, the constant term is $\sum_{{\bm i}\in \mathbb{P}_d}c_j\prod_{k=j}^{d}(-a_{i_k,j})$. Then, the constant term of $\sum_{j=1}^{d}S_j$ is $\prod_{i=1}^{d}(-a_{i,1})+
	\sum_{j=2}^{d}  \sum_{{\bm i}\in \mathbb{P}_d}c_j\prod_{k=j}^{d}(-a_{i_k,j})$. From above all, equation (14) is true.
\end{proof}	

\noindent
{\bf Speed-up the client-side computation.}
Theorem 5 shows that the client   has to compute $\prod_{i=1}^{d}m_i$ as
\begin{equation}
\label{eqn:dec}
 \sum_{j=1}^{d}S_j-\prod_{i=1}^{d}(-a_{i,1})-
 \sum_{j=2}^{d}
 \sum_{{\bm i}\in \mathbb{P}_d}
 c_j\prod_{k=j}^{d}(-a_{i_k,j}).
\end{equation}
  The client's local computation  incurred by
  (\ref{eqn:dec}) may be large.
To speed-up the client-side computation, we can distribute the
  computations of most monomials in (\ref{eqn:dec}) to the servers.
  We observe that any term
  $a_{i_1,1}a_{i_2,2}\cdots a_{i_d,d}$ with
  $|\{i_1,i_2,\ldots,i_d\}|<d$ will be  computable by at least one of
  the $d$ servers. In our construction, we will distribute any such term to one
  of the servers that can compute it.
  On the other hand, the term  $a_{i_1,1}a_{i_2,2}\cdots a_{i_d,d}$
  is not computable by any of the $d$ servers if and only if
$\{i_1,i_2,\ldots,i_d\}$ is a permutation of the set $[d]$.
The client will be responsible to compute such terms.

\subsection{Basic Ideas for Constructing $d$S-VDCLED}

In our $d${\sf S-DCLED} scheme each server
performs a computation of degree $\leq d$ over its data.
 By  using the
homomorphic MACs of \cite{CF18} one can make such computations can be made verifiable
and therefore obtain a
 $d${\sf S-VDCLED} scheme.

\section{Performance Analysis}

In this section, we shall
 implement the proposed
  schemes and compare with
    \cite{CF14}.
As we are mostly interested in the practicality of all schemes, the comparisons
between all schemes
will be done in terms of the running time of the server-side computations, and the running time of
 the client-side computations.
The comparison will be done with three experiments.
The first experiment will compare  the 2{\sf S-DCLED}/2{\sf S-VDCLED} from Section 4 with the   2{\sf S-DCED} from \cite{CF14}.
The second experiment will do the same comparisons   but in a scenario where a large number of
computation requests occur at the same time.

\subsection{Experiments Designs}

 We implement  all of the schemes with a security parameter $\lambda=128$ and  in a Ubuntu 16.04 LTS 64-bit operating
 system with 4GB RAM and Intel$^\circledR$ Core$^\circledR$ i7-6700 3.40GHz processor.
We choose  the PRF $F$ in all schemes as the  standard AES with 128-bit secret key from the library
   OpenSSL 1.0.2g.
   We choose the efficient Paillier
 cryptosystem \cite{Pai}, whose ciphertext size is  half of \cite{Pai99} and has a fast decryption algorithm,  as the   linearly homomorphic encryption  for
 the 2{\sf S-DCED} scheme from \cite{CF14}. We realize all   large integer  related mathematical computations   based on the C
   libraries GMP and FLINT.

In the  first experiment,  we consider the computation of a quadratic function
 $f(m_1,\dots,m_n)=\sum_{i,j\in[n]}\alpha_{i,j}m_im_j+\sum_{k\in[n]}\beta_{k}m_k+\gamma$
 on  outsourced data, where the number  $n$ of data items is chosen from
 $ \{10,50,100,500,1000\}$.
In the second experiment, we
choose   $n=500$ and consider $t$ simultaneous computation requests for
    $t\in\{10,100,1000,10000\}$. We compare
     between  \textsf{2S-DCED}, \textsf{2S-DCLED} and \textsf{2S-VDCLED} with
the  average  waiting time.

\begin{figure}[htp]
	\centering
	\includegraphics[width=3.5in]{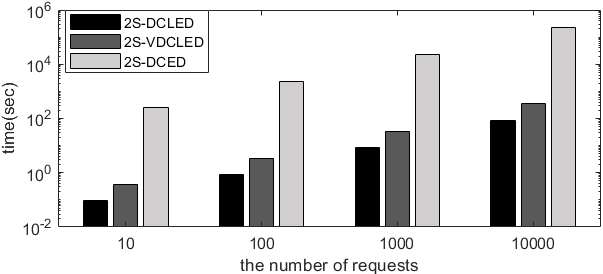}
	\caption{Average waiting time of \textsf{2S-DCED}, \textsf{2S-DCLED} and \textsf{2S-VDCLED}.}
	\centering
	\label{fig_sim}
\end{figure}

\subsection{Experimental Results}
Table 1 shows the evaluation algorithm and decryption algorithm execution times of \textsf{2S-DCED},
\textsf{2S-DCLED} and \textsf{2S-VDCLED}, where evaluation algorithm execution times
is the sum of the $\mathsf{2S.Eval_1}$ and $\mathsf{2S.Eval_2}$. Fig. 3 shows the
average waiting time  of the client for different number of requests.

\subsection{Comparisons}

We mainly compare the running time of the \textsf{2S-DCED} and our schemes \textsf{2S-DCLED} and \textsf{2S-VDCLED}  on the server-side and the client-side.

\vspace{1mm}
\noindent
{\bf Server-Side.} The servers perform the evaluation algorithms. For \textsf{2S-DCED}, the server-side need to run Paillier cryptosystem and large integer multiplication and exponentiations. For our schemes, the server-side need do multiplications and additions
 modulo $p$. Theoretically, in the server-side, our schemes are faster than \textsf{2S-DCED}. And our first experiment also confirmed it. More precisely, the  server-side running time of \textsf{2S-DCLED} and \textsf{2S-VDCLED} can be 2200 and 600 times faster than {\sf 2S-DCED} respectively.

\vspace{1mm}
\noindent
{\bf Client-Side.} The client perform the decryption algorithm.  The client-side running time of \textsf{2S-DCED} is dominated by the decryption of one Paillier ciphertext. For our schemes, the running time is dominated by the computation of $f(b_1,\dots,b_n)$. Then, the client-side running time of \textsf{2S-DCED} is fixed,  the client-side running time of our schemes become longer as the amount of data increases. But this is not a disadvantage.  When the amount of data is bounded, the running time of our schemes will be shorter than \textsf{2S-DCLED}. Even if the  client-side running time of our schemes are larger  than \textsf{2S-DCED}, their difference  is small. When $n=1000$, \textsf{2S-DCED}  is only 0.04 second faster than \textsf{2S-VDCLED}, which does not have obvious advantages in practical applications. Our second experiment shows that this advantage is not significant. In the second experiment, if the client receives multiple requests at the same time,  the time that \textsf{2S-DCED} responds to each request will be significantly higher than our schemes. This means that the long waiting time  would discourage the client from actually using the service and the numerous computations would  require more  charge by the client.

\section{Concluding Remarks}

In this paper, we proposed a multi-server model for delegating  computations on label-encrypted data.
We constructed  both a 2{\sf S-DCLED} scheme and
a 2{\sf S-VDCLED} scheme.
The server-side computations in both schemes are much faster than
the 2{\sf S-DCED} scheme from \cite{CF14}.
The client-side computations in both schemes are faster than
 \cite{CF14} when the size of the data is moderate.
 The semantic security of both schemes only depends on the mild assumption that PRFs exist.
The 2{\sf S-VDCLED} scheme also achieves verifiability, which was  not provided in \cite{CF14}.
We also extend the study to $d$-server schemes, which can delegate degree-$d$ computations,
a functionality not provided in \cite{CF14}.
The complexity of our decryption algorithm
depends on the size of the outsourced data.
Removing or weakening this dependency is an interesting open problem
for future work.

\section{Acknowledgments}
This work was supported by NSFC (No. 61602304) and Pujiang Talent Program (No. 16PJ1406500).

\bibliographystyle{plain}

\newcounter{mytempeqncnt}

\ifCLASSOPTIONcaptionsoff
  \newpage
\fi

%
%
%
%

\end{document}